\documentclass[10pt,twocolumn,twoside]{IEEEtran}
\usepackage{graphicx}
\UseRawInputEncoding
\usepackage{amsmath}
\usepackage{amsfonts}
\usepackage{array}
\newcommand{\PreserveBackslash}[1]{\let\temp=\\#1\let\\=\temp}
\newcolumntype{C}[1]{>{\PreserveBackslash\centering}p{#1}}
\newcolumntype{R}[1]{>{\PreserveBackslash\raggedleft}p{#1}}
\newcolumntype{L}[1]{>{\PreserveBackslash\raggedright}p{#1}}
\usepackage[usenames]{color}
\usepackage{colortbl,booktabs}
\usepackage{tabularx}
\usepackage{multicol}
\usepackage{booktabs}
\usepackage[Symbol]{upgreek}
\usepackage{subfigure}
\usepackage{bm}
\usepackage{color}
\usepackage{cite}
\usepackage{balance}
\usepackage[ruled,vlined]{algorithm2e}

\usepackage{threeparttable}
\usepackage{amsmath}
\usepackage{dcolumn}
\usepackage{multirow}
\usepackage{amsfonts}

\newcolumntype{d}[1]{D{.}{.}{#1}}

\begin{document}

\bibliographystyle{IEEEtran} 
\title{Deep Learning for Beamspace Channel \\Estimation in Millimeter-Wave Massive MIMO Systems}

\author{Xiuhong Wei, Chen Hu, and Linglong Dai

\thanks{This work has been accepted by IEEE Transactions on Communications.}
\thanks{All authors are with the Beijing National Research Center for Information Science and Technology (BNRist) as well as the Department of Electronic Engineering, Tsinghua University, Beijing 100084, China (e-mails: weixh19@mails.tsinghua.edu.cn, huc16@mails.tsinghua.edu.cn, daill@tsinghua.edu.cn).}
\thanks{This work was supported by the National Key Research and Development Program of China under Grant 2020YFB1805005 and the National Natural Science Foundation of China for Outstanding Young Scholars under Grant 61722109.}}

\maketitle
\vspace{-3em}
\begin{abstract}
Millimeter-wave massive multiple-input multiple-output (MIMO) can use a lens antenna array to considerably reduce the number of radio frequency (RF) chains, but channel estimation is challenging due to the number of RF chains is much smaller than that of antennas. By exploiting the sparsity of beamspace channels, the beamspace channel estimation can be formulated as a sparse signal recovery problem, which can be solved by the classical iterative algorithm named approximate message passing (AMP), and its corresponding version learned AMP (LAMP) realized by a deep neural network (DNN). However, these existing schemes cannot achieve satisfactory estimation accuracy. To improve the channel estimation performance, we propose a prior-aided Gaussian mixture LAMP (GM-LAMP) based beamspace channel estimation scheme in this paper. Specifically, based on the prior information that beamspace channel elements can be modeled by the Gaussian mixture distribution, we first derive a new shrinkage function to refine the AMP algorithm. Then, by replacing the original shrinkage function in the LAMP network with the derived Gaussian mixture shrinkage function, a prior-aided GM-LAMP network is developed to estimate the beamspace channel more accurately. Simulation results by using both the theoretical channel model and the ray-tracing based channel dataset show that, the proposed GM-LAMP network can achieve better channel estimation accuracy.
\end{abstract}

\section{Introduction}\label{S1}
Millimeter-wave (mmWave) massive multiple-input multiple-output (MIMO) has been considered as a key technique for 5G and beyond~\cite{5G}. In order to reduce the hardware cost and power consumption caused by a large number of antennas and the associated radio frequency (RF) chains, the lens antenna array has been recently investigated to provide an energy-efficient realization of hybrid precoding for mmWave massive MIMO~\cite{Ai,zeng16mmwave}. By employing the lens antenna array, which can concentrate signals from different directions on different antennas, the spatial channel can be converted to the beamspace channel~\cite{Zeng_2014}. As there are only a few dominant propagation paths with large path gains at mmWave frequencies, the beamspace channel in mmWave massive MIMO systems is sparse in nature~\cite{brady2013beamspace}. Therefore, by only selecting a small number of dominant beams, the number of RF chains connected to the digital baseband can be considerably reduced. Beam selection requires the accurate channel state information (CSI) in the beamspace~\cite{srinidhi11}, which is challenging due to the high channel dimension, especially when the number of RF chains is much smaller than the number of antennas~\cite{sayeed2013beamspace,amadorilow,gao16bs}.

\subsection{Prior works}\label{S1.1}
There are some recently proposed schemes for beamspace channel estimation. Specifically,~\cite{ZengCE} proposed a two-way channel estimation scheme with low computational complexity, where the antennas corresponding to the dominant beams are firstly determined by beam training between the base station (BS) and users, and then only channel elements corresponding to these selected antennas are estimated. However, the number of pilot symbols required to scan all possible beams is proportional to the number of BS antennas, which is very large (e.g., 256 antennas). Furthermore, by exploiting the sparsity of beamspace channels, some classical compressive sensing (CS) based schemes could estimate the beamspace channel with a reduced pilot overhead~\cite{alkhateeb2014channel,HuangCE2,YuenCE}, such as the orthogonal matching pursuit (OMP) algorithm used in~\cite{alkhateeb2014channel}. Apart from the sparsity, the beamspace channel may exhibit angular spreads. Based on this channel characteristics,~\cite{FangCE} proposed a two-stage CS method for channel estimation, which consists of a matrix completion stage and a sparse recovery stage.

Unfortunately, all of these beamspace channel estimation schemes above~\cite{alkhateeb2014channel,HuangCE2,YuenCE,FangCE} cannot achieve satisfactory estimation accuracy in low signal-to-noise ratio (SNR) regions, and they also have high computational complexity especially when the sparsity level of the beamspace channel is high. As a powerful iterative algorithm for sparse signal recovery, the approximate message passing (AMP) algorithm can be used to estimate the beamspace channel with low computational complexity~\cite{AMP1,FangMP}. However, it is difficult to find the optimal shrinkage parameters for the AMP algorithm (the empirical shrinkage parameters are usually used instead), which restricts its channel estimation performance in practice.

Recently, the amazing success of deep learning (DL) in other fields like image recognition~\cite{IM1,IM2} and speech processing~\cite{Speech} has greatly inspired researchers to use this powerful tool to solve some problems in wireless communications~\cite{DL2,DL3,DL5,BDL}. With the help of DL, we can extract underlying features of wireless big data and provide some improved solutions to some complicated problems in wireless communications, such as low density parity check (LDPC) decoding~\cite{DL2}, sparse code multiple access (SCMA) codebook design~\cite{DL3}, end-to-end communications~\cite{DL5}, and hybrid precoding for massive MIMO~\cite{BDL}.

Inspired by the powerful learning ability of deep neural networks (DNNs), some DL based channel estimation schemes have been proposed.~\cite{DDL} was the first work to exploit the DL tool for channel estimation in the wireless energy transfer system. They developed an autoencoder based channel estimation scheme, where the encoder is used to design pilots and the decoder is used to estimate channels. In order to make the pilot length smaller than the number of the antennas in the massive MIMO system,~\cite{ADL} proposed a joint pilot and data aided channel estimation scheme using DNNs, where the pilot-aided estimation process is realized by a two-layer neural network together with a DNN and the data-aided estimation process is realized by another DNN.~\cite{CDL} proposed a DL based super-resolution channel estimation scheme in the mmWave massive MIMO system, which leverages a DNN for direction-of-arrival (DOA) estimation.

All of these works above~\cite{DDL,ADL,CDL} adopt classical DNNs (e.g., multilayer perceptron) to solve the channel estimation problems and achieve better performance in different scenarios. However, their neural networks are usually regarded as the black-boxes, which may lack stable performance guarantees~\cite{DeepUnfolding}. By contrast, there are some other DL based channel estimation schemes by integrating the conventional algorithms that have certain performance guarantees with the DL tool.~\cite{LDAMP} has presented a learned denoising-based approximate message passing (LDAMP) network for channel estimation, where a denoising convolutional neural network (DnCNN) for image recovery is incorporated into the AMP algorithm to replace the original shrinkage function. In order to improve the performance of the sparse signal recovery,~\cite{LAMP} proposed a learned AMP (LAMP) network by directly unfolding the iterations of the AMP algorithm into the corresponding layer-wise network structure, where its linear transform coefficients and nonlinear shrinkage parameters are jointly optimized by the DNN. However, when used to solve the beamspace channel estimation problem, the existing LAMP network cannot achieve the satisfactory estimation accuracy.

\subsection{Our contributions}\label{S1.2}
In this paper, in order to improve the estimation performance, we propose a complex-valued Gaussian mixture LAMP (GM-LAMP) based beamspace channel estimation scheme by fully utilizing the prior information of the beamspace channel$^1$.

Specifically, we first exploit the prior information that beamspace channel elements follow the Gaussian mixture distribution to derive a new shrinkage function. Then, by replacing the original shrinkage function in the LAMP network with the derived Gaussian mixture shrinkage function, a prior-aided GM-LAMP network is developed to estimate the beamspace channel. Finally, we verify our work by using the widely used channel model for theoretical analysis and the publicly-available channel dataset based on ray-tracing, respectively. Simulation results show that, compared with conventional algorithms, the proposed GM-LAMP network can achieve better estimation accuracy in the above two channels.

\footnotetext[1]{
Simulation codes are provided to reproduce the results presented in this paper: http://oa.ee.tsinghua.edu.cn/dailinglong/publications/publications.html.
}

\subsection{Organization and notation}\label{S1.3}
The rest of the paper is organized as follows. In section II, the beamspace channel estimation problem in mmWave massive MIMO systems is formulated as a sparse signal recovery problem, and the conventional AMP algorithm and LAMP network for solving this problem are briefly reviewed. In section III, we derive the new shrinkage function based on the Gaussian mixture distribution, and propose the GM-LAMP network for improved beamspace channel estimation. The computational complexity of the proposed scheme is also analyzed in Section III. Simulation results are provided to show the performance of the proposed GM-LAMP network in Section IV. Finally, conclusions are given in Section V.

{\it Notation}: Lower-case and upper-case boldface letters ${\bf{a}}$ and ${\bf{A}}$ denote a vector and a matrix, respectively; $\bf{a}^{*}$ denotes the conjugate of vector $\bf{a}$; ${{{\bf{A}}^H}}$ and ${{{\bf{A}}^{T}}}$ denote the conjugate transpose and transpose of matrix ${\bf{A}}$, respectively; ${{\left\|  \bf{a}  \right\|_2}}$ denotes the ${{l_2}}$-norm of vector ${\bf{a}}$; ${\left|  a  \right|}$ denotes the amplitude of scalar ${a}$; ${a^*}$ denotes the conjugate of scalar ${a}$; ${\bf{A}}{\otimes}{\bf{B}}$ denotes the Kronecker product of ${\bf{A}}$ and ${\bf{B}}$; ${{\cal U}(-a,a)}$ denotes the probability density function of uniform distribution on $(-a,a)$; $\delta \left( x \right)$ denotes the Dirac delta function; $\rm{sinc}\left( x \right){\buildrel {\rm{\Delta}} \over =}\frac{\sin\left(N\pi{x}\right)}{N\pi{x}}$ denotes the Dirichlet sinc function. Finally, ${{{\mathbf{I}}_{K}}}$ is the ${K \times K}$ identity matrix.

\section{System Model}\label{S2}
In this section, we first introduce the beamspace channel model, and then formulate the beamspace channel estimation problem as a sparse signal recovery problem. Finally, the conventional AMP algorithm~\cite{AMP1} and its corresponding LAMP network proposed in~\cite{LAMP} to solve this problem are reviewed.

\subsection{Beamspace channel}\label{S2.1}
We consider a time division duplex (TDD) based mmWave massive MIMO system, as shown in Fig. 1~\cite{gao16beamspace}, where the BS employs a lens antenna array with ${N}$ antennas and ${{N_{{\rm{RF}}}}}$ RF chains to simultaneously serve ${K}$  single-antenna users.

\begin{figure}[tp]
\begin{center}
\includegraphics[width=1\linewidth]{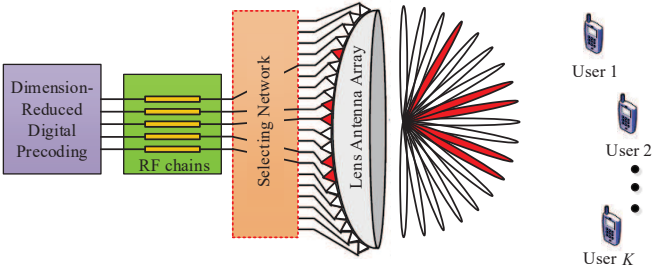}
\end{center}
\setlength{\abovecaptionskip}{-0.0cm}
\caption{mmWave massive MIMO with lens antenna array~\cite{gao16beamspace}.} \label{FIG1}
\end{figure}

In order to formulate the beamspace channel estimation problem, we start with the conventional mmWave massive MIMO channel in the spatial domain. According to the widely used Saleh-Valenzuela channel model~\cite{alkhateeb2014channel}, the channel vector ${\bf{h}}_k$ of size ${N \times 1}$  between the ${k}$th $\left({k = 1,2, \cdots ,K}\right)$ user and the $N$-antenna BS can be presented by
\begin{equation}\label{eq1}
{{\bf{h}}_k} = \sqrt {\frac{N}{{{L_k}}}} \sum\limits_{l = 1}^{{L_k}} {\beta _{k,l}}{\bf{a}}\left( {\theta_{k,l}^{\rm{azi}},{\theta_{k,l}^{\rm{ele}}}}\right)  = \sqrt {\frac{N}{{{L_k}}}} \sum\limits_{l = 1}^{{L_k}} {{{\bf{c}}_{k,l}}},
\end{equation}
where ${L_k}$ is the number of resolvable paths, and ${{{\bf{c}}_{k,l}}}={\beta _{k,l}}{\bf{a}}\left( {\theta_{k,l}^{\rm{azi}},{\theta_{k,l}^{\rm{ele}}}}\right)$ is the $l$th path component. ${\beta _{k,l}}$, ${\theta_{k,l}^{\rm{azi}}}$ and ${\theta_{k,l}^{\rm{ele}}}$ are the complex gain, azimuth angle and elevation angle of the ${l}$th path, respectively. ${\bf{a}}\left( {\theta_{k,l}^{\rm{azi}},{\theta_{k,l}^{\rm{ele}}}}\right)$ is the ${N \times 1}$ array steering vector, which depends on the array geometry. Ignoring the subscripts without loss of generality, for the simpler uniform linear arrays (ULAs), the array steering vector can be determined by one angle, which can be presented by~\cite{alkhateeb2014channel}
\begin{equation}\label{eq2}
{\bf{a}}_{\rm{ULA}}\left( \theta\right) = \frac{1}{{\sqrt N }}{\left[ {{e^{ - j2{\pi}d{\rm{sin}}\left(\theta\right)\bf{n}/{\lambda}}}} \right]},
\end{equation}
where ${\bf{n}}{\rm{ = }}{\left[ { 0, 1, \cdots ,N-1} \right]^T}$. For the widely considered uniform planar arrays (UPAs) with ${N_1 \times N_2}$ ($N=N_1 \times N_2$) antennas, we have~\cite{el2013spatially}
\begin{equation}\label{eq3}
\begin{split}
{\bf{a}}_{\rm{UPA}}\left( \theta^{\rm{azi}}, \theta^{\rm{ele}}  \right) = \frac{1}{{\sqrt N }}{\left[ {{e^{ - j2{\pi}d{\rm{sin}}\left(\theta^{\rm{azi}}\right){\rm{sin}}\left(\theta^{\rm{ele}}\right) {\bf{n}}_1/{\lambda}}}} \right]}\\{\otimes}{\left[ {{e^{ - j2{\pi}d{\rm{cos}}\left(\theta^{\rm{ele}}\right) {\bf{n}}_2/{\lambda}}}} \right]},
\end{split}
\end{equation}
where ${{\bf{n}}_1}={\left[ { 0, 1, \cdots ,N_1-1} \right]^T}$ and ${{\bf{n}}_2}={\left[ { 0, 1, \cdots ,N_2-1} \right]^T}$. In~(\ref{eq2}) and~(\ref{eq3}), ${\lambda }$ is the wavelength of carrier, and ${d}$ is the antenna spacing usually satisfying ${d = \lambda /2}$ in mmWave communications~\cite{han2015large}. Then, we can respectively define $\psi  \buildrel \Delta \over = {d}{\rm{sin}}\left({\theta}\right)/{\lambda }$ as the spatial angle for ULAs, and ${\psi}^{\rm{azi}}\buildrel \Delta \over = {d}{\rm{sin}}\left({\theta^{\rm{azi}}}\right){\rm{sin}}\left({\theta^{\rm{ele}}}\right)/{\lambda }$ and $\psi^{\rm{ele}}  \buildrel \Delta \over = {d}{\rm{cos}}\left({\theta^{\rm{ele}}}\right)/{\lambda }$ as the spatial angles for UPAs.

The spatial domain channel can be directly transformed to the beamspace channel by using a lens antenna array. As a matter of fact, the lens antenna array plays the role of a spatial discrete fourier transform (DFT) matrix ${{\bf{U}}}$ of size ${N \times N}$~\cite{gao16beamspace}. For ULAs, the matrix ${{\bf{U}}}$ can be expressed as
\begin{equation}\label{eq4}
{\bf{U}} = {\left[ {{\bf{\bar a}}_{\rm{ULA}}\left( {{{\bar \psi }_1}} \right),{\bf{\bar a}}_{\rm{ULA}}\left( {{{\bar \psi }_2}} \right), \cdots ,{\bf{\bar a}}_{\rm{ULA}}\left( {{{\bar \psi }_N}} \right)} \right]^H},
\end{equation}
where ${{\bar \psi _n} = \frac{1}{N}\left( {n - \frac{{N + 1}}{2}} \right)}$ for ${n = 1,2, \cdots ,N}$ are the spatial directions predefined by the lens antenna array. Similar to~(\ref{eq2}), ${\bf{\bar a}}_{\rm{ULA}}\left( {{{ \psi }}} \right)$ can be presented by
\begin{equation}\label{eq5}
{\bf{\bar a}}_{\rm{ULA}}\left( {\psi}\right) = \frac{1}{{\sqrt N }}{\left[ {{e^{ - j2{\pi}{\psi}\bf{n}}}} \right]}.
\end{equation}
For UPAs, ${{\bf{U}}}$ can be expressed as
\begin{equation}\label{eq6}
\begin{aligned}
{\bf{U}} =\left[ {\bf{\bar a}}_{\rm{UPA}}\left(\bar \psi^{\rm{azi}}_1,\bar \psi^{\rm{ele}}_1  \right), \cdots ,{\bf{\bar a}}_{\rm{UPA}}\left(\bar \psi^{\rm{azi}}_1,\bar \psi^{\rm{ele}}_{N_2}\right),\cdots, \right.\\\left.{\bf{\bar a}}_{\rm{UPA}}\left(\bar \psi^{\rm{azi}}_{N_1},\bar \psi^{\rm{ele}}_{1}  \right), \cdots ,{\bf{\bar a}}_{\rm{UPA}}\left(\bar \psi^{\rm{azi}}_{N_1},\bar \psi^{\rm{ele}}_{N_2}  \right)  \right]^H,
\end{aligned}
\end{equation}
where ${{\bar \psi^{\rm{azi}}_n} = \frac{1}{N_1}\left( {n - \frac{{N_1 + 1}}{2}} \right)}$ for ${n = 1,2, \cdots ,N_{1}}$ and ${{\bar \psi^{\rm{ele}}_n} = \frac{1}{N_2}\left( {n - \frac{{N_2 + 1}}{2}} \right)}$ for ${n = 1,2, \cdots ,N_{2}}$ are respectively predefined spatial angles of azimuth and elevation by the lens antenna array. Similar to~(\ref{eq3}), ${\bf{\bar a}}_{\rm{UPA}}\left(\psi^{\rm{azi}},\psi^{\rm{ele}}  \right)$ can be presented by
\begin{equation}\label{eq7}
{\bf{\bar a}}_{\rm{UPA}}\left(\psi^{\rm{azi}}, \psi^{\rm{ele}}  \right) = \frac{1}{{\sqrt N }}{\left[ {{e^{ - j2{\pi}{{ \psi^{\rm{azi}}}{\bf{n}}_1}}}} \right]{\otimes}\left[ {{e^{ - j2{\pi}{{ \psi^{\rm{ele}}}{\bf{n}}_2}}}} \right]}.
\end{equation}

Finally, the beamspace channel vector ${{\tilde{\bf{{h}}}_k}}$ of size ${N \times 1}$  between the ${k}$th user and the $N$-antenna BS can be presented by
\begin{equation}\label{eq8}
{{\tilde{\bf{{h}}}_k}} = {\bf{U}}{\bf{h}}_k = \sqrt {\frac{N}{{{L_k}}}} \sum\limits_{l = 1}^{{L_k}} {{{\tilde {\bf{c}}}_{_{k,l}}}},
\end{equation}
where ${{\tilde {{\bf{c}}}_{k,l}} = {\bf{U}}{{\bf{c}}_{k,l}}}$  is the ${l}$th channel component of the beamspace channel.
\subsection{Problem formulation }\label{S2.2}
In order to acquire the CSI, all users should transmit the known pilot symbols to the BS over ${Q}$ instants. Due to the reciprocity of the TDD channel, we can only consider the uplink to formulate the channel estimation problem. Then, the downlink channel can be directly obtained according to the estimated uplink channel. In this paper, we adopt the widely used orthogonal pilot transmission strategy~\cite{gao16beamspace}, where the uplink channel estimation for each user is independent due to the pilot orthogonality, and thus we can estimate the beamspace channel vectors between all $K$ users and the BS one by one. Without loss of generality, we take the beamspace channel vector ${{\tilde{\bf{{h}}}_k}}$ between the ${k}$th user and the BS as an example to formulate the channel estimation problem.

In the ${q}$th instant for pilot transmission, the ${{N_{\rm{RF}}} \times 1}$ measurement signal ${{\bf{y}}_{k,q}}$ in the baseband at the BS after beam selection can be presented as~\cite{gao16beamspace}
\begin{equation}\label{eq9}
{{\bf{y}}_{k,q}} = {{\bf{A}}_{k,q}{{\tilde{\bf{{h}}}_k}}{s_{k,q}} + {\bf{\bar n}}_{k,q}}, q = 1,2, \cdots ,Q,
\end{equation}
where ${{\bf{A}}_{k,q}}$ is the ${{N_{\rm{RF}}} \times N}$ beam selection network, ${s_{k,q}}$ is the transmitted pilot symbol, ${\bf{\bar n}}_{k,q} = {{{\bf{A}}_{k,q}}{\bf{n}}}_{k,q}$ is the effective noise vector, where ${{\bf{n}}_{k,q}}\sim{\cal C}{\cal N}\left( {0,\sigma _n^2{\bf{I}_{N}}} \right)$ is the ${{N} \times 1}$ noise vector with ${\sigma _n^2}$ representing the noise power.

After ${Q}$ instants of pilot transmission, we can obtain the ${{M \times 1}\left( {M = Q{N_{\rm{RF}}}} \right)}$ overall measurement signal ${{\bf{y}}_k}$ by assuming ${s_{k,q}} = 1$ for ${q = 1,2, \cdots ,Q}$ as
\begin{equation}\label{eq10}
{{\bf{y}}_k} = \left[ {\begin{array}{*{20}{c}}
{{{\bf{y}}_{k,1}}}\\
{{{\bf{y}}_{k,2}}}\\
 \vdots \\
{{{\bf{y}}_{k,Q}}}
\end{array}} \right] = {{\bf{A}}_k}{{\tilde{\bf{{h}}}_k}} + {{\bf{n}}_k},
\end{equation}
where ${\bf{A}}_k = {[{\bf{A}}_{k,1}^T,{\bf{A}}_{k,2}^T, \cdots ,{\bf{A}}_{k,Q}^T]^T}$ is the ${M \times N}$ selection matrix with the entry being $\pm \frac{1}{{\sqrt {M} }}$~\cite{gao16beamspace}, and ${\bf{n}}_k = {[{\bf{\bar n}}_{k,1}^T,{\bf{\bar n}}_{k,2}^T, \cdots ,{\bf{\bar n}}_{k,Q}^T]^T}$ is the ${M \times 1}$ effective noise vector for ${Q}$ instants.

Since the channel estimation method is the same for all $K$ users due to the pilot orthogonality, the subscript $k$ in the problem~(\ref{eq10}) can be omitted, then~(\ref{eq10}) can be expressed as
\begin{equation}\label{eq11}
{\bf{y}} = {\bf{A}}{{\tilde{\bf{{h}}}}} + {\bf{n}}.
\end{equation}
Note that only the elements of the beamspace channel ${{\tilde{\bf{{h}}}}}$ that are close to the practical spatial angles of the channel paths have large values. As there are only a few propagation paths due to limited scattering at mmWave frequencies, the beamspace channel ${{\tilde{\bf{{h}}}}}$ is approximately sparse~\cite{brady2013beamspace}. Consequently, we can apply the sparse signal recovery algorithms in CS to estimate the beamspace channel with a low pilot overhead, where the matrix ${\bf{A}}$ in~(\ref{eq11}) can be regarded as the sensing matrix in CS. That is to say, the beamspace channel estimation problem in~(\ref{eq11}) can be formulated as a sparse signal recovery problem
\begin{equation}\label{eq12}
\min {\left\| {{{\tilde{\bf{{h}}}}}}  \right\|_0},\quad\quad{\rm{s.t.}}{\rm{ }}{\left\| {{\bf{y}} - {\bf{A}} {{{\tilde{\bf{{h}}}}}}} \right\|_2} \le \varepsilon,
\end{equation}
where ${\left\| {{{\tilde{\bf{{h}}}}}}  \right\|_0}$ is the number of non-zero elements of ${{{\tilde{\bf{{h}}}}}}$, ${\varepsilon}$ is the error tolerance parameter.

Due to the non-convexity of the $l_0$-norm, the problem in~(\ref{eq12}) is NP-hard~\cite{B1995}. Therefore, this problem is usually converted to a convex optimization problem by replacing the $l_0$-norm with the $l_1$-norm~\cite{J2002,JA2004,D2006}. There have been some conventional greedy algorithms to solve it, such as OMP~\cite{alkhateeb2014channel} and compressive sampling matching pursuit (CoSaMP)~\cite{CoSaMP}. However, these greedy algorithms cannot achieve satisfactory estimation accuracy. Especially, with the increase of the sparsity level, the computational complexity of the greedy algorithms will be also increased.

\subsection{AMP algorithm and LAMP network}\label{S2.3}
Since the number of antennas in mmWave massive MIMO systems is usually large, the dimension of the sparse signal in~(\ref{eq12}) is high. Thanks to faster convergence, the iterative AMP algorithm can be used to recover the sparse signal with low computational complexity, especially for the high-dimensional sparse signal~\cite{AMP1}. In this subsection, we introduce how the complex-valued AMP algorithm estimates the beamspace channel, as shown in \textbf{Algorithm 1}.

\begin{algorithm}[h]
\caption{Approximate Message Passing (AMP)}
\KwIn{The measurement vector ${{\bf{y}}}$, the sensing matrix ${{\bf{A}}}$, the number of iterations ${T}$.}
\textbf{Initialization}: ${{\bf{v}}_{-1}} = {\bf{0}},{b_0} = 0,{c_0} = 0,{{{\hat{\tilde{\bf{{h}}}}}_{0}}} = {\bf{0}}$.
\\\textbf{for} $t = 0,1,\cdots,T-1$ \textbf{do}
 \\1. \hspace*{+1mm} ${{\bf{v}}_t} = {\bf{y}} - {\bf{A}}{{\hat{\tilde{\bf{{h}}}}}_{t}} + {b_{t}}{{\bf{v}}_{t - 1}}+{c_{t}}{{\bf{v}}^{*}_{t - 1}}$
 \\2. \hspace*{+1mm} $\sigma _t^2 = \frac{1}{M}\left\| {{{\bf{v}}_t}} \right\|_2^2$
 \\3. \hspace*{+1mm} ${\bf{r}}_t = {{\hat{\tilde{\bf{{h}}}}}_{t}} + {{\bf{A}}^T}{{\bf{v}}_t}$
 \\4. \hspace*{+1mm} ${{\hat{\tilde{\bf{{h}}}}}_{t+1}} = {{\bm{\eta}} _{\rm{st}}}({{\bf{r}}_t};{\lambda _t},\sigma _t^2)$
 \\5. \hspace*{+1mm} ${b_{t+1}} = \frac{1}{M}\sum\limits_{i=1}^{N}\frac{{\partial{{{\eta}}_{\rm{st}}\left( {{{{r}}_{t,i}};{\lambda _t},\sigma_t^2}\right)}}}{\partial{{{r}}_{t,i}}}$
 \\6. \hspace*{+1mm} ${c_{t+1}} = \frac{1}{M}\sum\limits_{i=1}^{N}\frac{{\partial{{{\eta}}_{\rm{st}}\left( {{{{r}}_{t,i}};{\lambda _t},\sigma_t^2}\right)}}}{\partial{{{r}}^{*}_{t,i}}}$
 \\\textbf{end for}
\\\KwOut{Sparse signal recovery results: ${{\hat{\tilde{\bf{{h}}}}}}={{\hat{\tilde{\bf{{h}}}}}_{T}}$.}
\end{algorithm}

In \textbf{Algorithm 1}, the term ${b_{t}}{\bf{v}}_{t - 1}$ and term ${c_{t}}{\bf{v}}^{*}_{t - 1}$ in Step 1 are called Onsager Correction~\cite{AMP1}, which are introduced into the AMP algorithm to accelerate the convergence. The critical step of the AMP algorithm is Step 4, in which the estimate ${\hat{\tilde{\bf{{h}}}}}_{t+1}$ in the $t$th iteration is obtained through the soft threshold shrinkage function ${{\bm{\eta}} _{\rm{st}}}$: ${\mathbb{C}^N \rightarrow \mathbb{C}^N}$. The shrinkage function ${\bm{\eta}} _{\rm{st}}$ is nonlinear element-wise operation, which takes the sparsity of the vector ${{\tilde{\bf{{h}}}}}$ into consideration, and makes the estimate ${\hat{\tilde{\bf{{h}}}}}_{t+1}$ sparser. For the $i$th element $r_{t,i}={\left| r_{t,i} \right|{e^{j{\omega}_{t,i} }}}$ (${i = 1,2, \cdots ,N}$) of input vector ${{\bf{r}}_{t}}$, we have
\begin{equation}\label{eq13}
\begin{aligned}
{\left[{{\bm{\eta}} _{\rm{st}}\left( {{{\bf{r}}_t};{\lambda _t},\sigma _t^2} \right)}\right]}_i
&={\eta _{\rm{st}}}\left( {\left| r_{t,i} \right|{e^{j{\omega}_{t,i} }};{\lambda _t},\sigma _t^2} \right)\\
&= \max \left( \left| r_{t,i} \right| - \lambda _t{\sigma _{t},0} \right){e^{j{\omega}_{t,i} }},
\end{aligned}
\end{equation}
where ${\omega}_{t,i}$ is the phase of complex-valued element $r_{t,i}$, ${\lambda _t}$ is the predefined and fixed parameter in the ${t}$th iteration, and $\sigma _t^2$ is updated via estimating the noise variance in Step 2. From~(\ref{eq13}), we can find that the soft threshold shrinkage function ${{\bm{\eta}} _{\rm{st}}}$ can shrink the amplitude of complex-valued input with low power to zero. In Step 5 and Step 6, the element-wise derivatives of the shrinkage function ${{\bm{\eta}} _{\rm{st}}}$ at the input vector $\bf{r}$ and its conjure vector ${\bf{r}}^{*}$ are respectively calculated to obtain $b_{t+1}$ and $c_{t+1}$.

Although the AMP algorithm is good at dealing with the large-scale sparse signal recovery problem, there are still two problems when it is used for the sparse beamspace channel estimation. First, the shrinkage parameter ${\lambda _t}$ in~(\ref{eq13}) usually takes the same empirical value for all iterations. Second, the general AMP algorithm cannot fully exploit the prior distribution of the beamspace channel. These two problems limit the performance of the AMP algorithm.
\begin{figure}[hbtp]
\begin{center}
\vspace*{0mm}\includegraphics[width=1\linewidth]{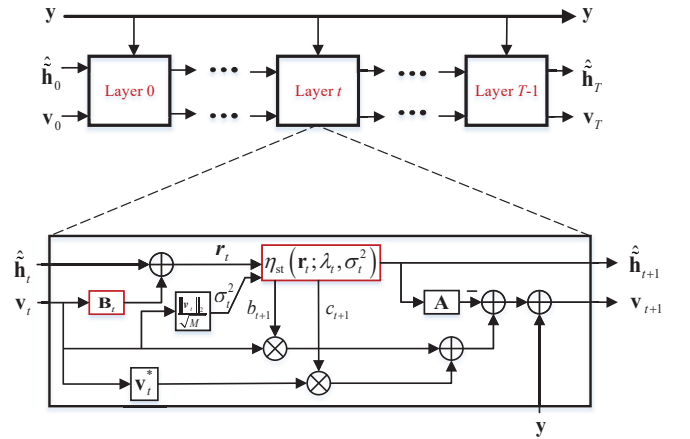}
\end{center}
\setlength{\abovecaptionskip}{-0.0cm}
\vspace*{0mm}\caption{LAMP network structure (the $t$th layer is explained in detail)~\cite{LAMP}.} \label{FIG2}
\end{figure}

To solve the first problem, the LAMP network based on the classical AMP algorithm has been recently proposed to optimize the nonlinear shrinkage parameter ${\lambda _t}$ in each iteration~\cite{LAMP}. As shown in Fig. 2, each iteration of the classical AMP algorithm is mapped to each layer of the LAMP network. To be specific, the inputs of the $t$th layer are ${\bf{y}} \in {\mathbb{C}^M}$, ${{\hat{\tilde{\bf{{h}}}}}_{t}} \in {\mathbb{C}^N}$ and ${\bf{v}}_{t} \in {\mathbb{C}^M}$, where ${\bf{y}}$ is the measurement vector in~(\ref{eq11}), ${\hat{\tilde{\bf{{h}}}}}_{t}$ and ${\bf{v}}_{t}$ are the outputs of the $\left(t-1\right)$th layer. Following the principle of the AMP algorithm, each layer of the LAMP network processes the signal as follows, which is similar to \textbf{Algorithm 1}:
\begin{equation}\label{eq14}
{{\hat{\tilde{\bf{{h}}}}}_{t+1}} = {{\bm{\eta}}_{\rm{st}}}({{\bf{r}}_t};{\lambda _t},\sigma _t^2),
\end{equation}
\begin{equation}\label{eq15}
{{\bf{v}}_{t+1}} = {\bf{y}} - {\bf{A}}{{\hat{\tilde{\bf{{h}}}}}_{t}} + {b_{t+1}}{{\bf{v}}_{t}}+{c_{t+1}}{{\bf{v}}_{t}^{*}},
\end{equation}
where
\begin{equation}\label{eq16}
{{\bf{r}}_t} = {{\hat{\tilde{\bf{{h}}}}}_{t}} + {{\bf{B}}_t}{{\bf{v}}_t},
\end{equation}
\begin{equation}\label{eq17}
\sigma _t^2 = \frac{1}{M}\left\| {{{\bf{v}}_t}} \right\|_2^2,
\end{equation}
\begin{equation}\label{eq18}
{b_{t+1}}=\frac{1}{M}\sum\limits_{i=1}^{N}\frac{{\partial{{{\eta}}_{\rm{st}}\left({{{{r}}_{t,i}};{\lambda_t},\sigma_t^2}\right)}}}{\partial{{{r}}_{t,i}}},
\end{equation}
\begin{equation}\label{eq19}
{c_{t+1}} = \frac{1}{M}\sum\limits_{i=1}^{N}\frac{{\partial{{{\eta}}_{\rm{st}}\left( {{{{r}}_{t,i}};{\lambda _t},\sigma_t^2}\right)}}}{\partial{{{r}}^{*}_{t,i}}},
\end{equation}where the shrinkage function ${{\bm{\eta}}_{\rm{st}}}$ of the AMP algorithm plays a role of the nonlinear activation function in the conventional DNN~\cite{LAMP}. What's more, from~(\ref{eq16}), we can find that different from the Step 3 in \textbf{Algorithm 1}, the LAMP network can choose the different linear coefficients ${{\bf{B}}_t}$ for each layer $t$, which can replace ${\bf{A}}^{T}$ as the linear transform from the measurement signal space to the original sparse signal space. It is worth noting that ${\bf{A}}^{T}$ is selected only for the convenience of derivation in the AMP algorithm. In the training stage of the LAMP network, the linear transform coefficients ${{\bf{B}}_t}$ of size ${N\times M}$ in~(\ref{eq16}) and the nonlinear shrinkage parameters ${\lambda _t}$ in~(\ref{eq14}),~(\ref{eq18}) and~(\ref{eq19}) can be optimized. Therefore, given enough training data, the LAMP network can find better shrinkage parameters by leveraging the powerful learning ability of the DNN.

However, the second problem of the AMP algorithm for beamspace channel estimation has not been solved. The conventional AMP algorithm and its corresponding LAMP network only consider the sparsity of signals to be recovered, which are general for any sparse signal recovery problem. In particular, compared with the activation function without an explicit physical meaning in the conventional DNN, the shrinkage function of the LAMP network is not specifically designed for the beamspace channel estimation problem under investigation. The LAMP based beamspace channel estimation schemes still cannot achieve satisfactory estimation accuracy. In order to improve the estimation accuracy, we will utilize the prior distribution of the sparse beamspace channel to propose a more suitable network for the beamspace channel estimation problem in mmWave massive MIMO systems in the next section.

\section{Proposed GM-LAMP Network for Beamspace Channel Estimation}\label{S3}
In this section, we first derive a new shrinkage function according to the Gaussian mixture distribution of beamspace channel elements. Then, based on the derived shrinkage function, the GM-LAMP based beamspace channel estimation scheme is proposed. After that, we also discuss how to extend the idea of the GM-LAMP network for other sparse signal recovery problems. Finally, the computational complexity analysis between the proposed algorithm and the existing algorithms is provided.

\subsection{Gaussian mixture distribution and its corresponding shrinkage function}\label{S3.1}
As we all know, we are likely to get a more accurate estimate with more prior information of the channel. Next, we will utilize more specific prior distribution (besides sparsity) of the beamspace channel to refine the LAMP network.

There have been some previous works to consider the Gaussian mixture distribution to model the prior distribution of beamspace channel elements for ULAs~\cite{YuenCE} and for UPAs~\cite{GM2} and verify its validity. Specifically, the probability density function of the element ${\tilde{h}}$ of the beamspace channel ${\tilde{\bf{h}}}$ can be expressed as:
\begin{equation}\label{eq20}
p\left( {\tilde{h}};{\bm{\theta} } \right) = \sum\limits_{k = 0}^{{N_c} - 1} {{p_k}{\cal C}{\cal N}\left( {\tilde{h}};\mu_k,\sigma _k^2 \right)},
\end{equation}where ${\bm{\theta}} = \left\{ {{p_0}, \cdots ,{p_{{N_c} - 1}},\mu_0,\cdots,\mu_{N_c-1},\sigma _0^2, \cdots ,\sigma _{{N_c} - 1}^2} \right\}$ is the set of all distribution parameters. ${N_c}$ is the number of Gaussian components in the Gaussian mixture distribution, ${p_k}$ is the probability of $k$th Gaussian component, $\mu_k$ and $\sigma _k^2$ denote the mean and variance of the $k$th Gaussian component, respectively. ${{\cal CN}\left({\tilde{h}}; \mu_k,\sigma_k^2 \right)=\frac{\rm{1}}{{\pi {\sigma_k^{\rm{2}}}}}{{\rm{e}}^{ - \frac{{{\left({\tilde{h}}-{\mu_k}\right)}^{*}}{{\left({\tilde{h}}-{\mu_k}\right)}}}{\sigma_k^{{\rm{2}}}}}}}$ denotes the probability density function of the $k$th Gaussian component. Take the ULA as an example, the rationality of the Gaussian mixture distribution can be explained based on the following two observations.

From~(\ref{eq1}),~(\ref{eq4}) and~(\ref{eq8}), the ${n}$th element ${\tilde{h}_n}$ of the beamspace channel ${\tilde{\bf{h}}}$ can be expressed by
\begin{equation}\label{eq21}
{\tilde{h}_n}= \sqrt {\frac{N}{{{L}}}} \sum\limits_{l = 1}^{{L}} {\beta _{l}}{{\rm{sinc}}\left( \Delta\psi_n\right)},
\end{equation}
where $\Delta\psi_n={\bar \psi_n}-\psi_l$. Firstly, it is noted that the complex gain ${\beta _{l}}$ follows the complex Gaussian distribution. Secondly, when the practical spatial direction $\psi_l$ for the $l$th path is close to the predefined spatial direction ${\bar \psi_n}$, ${{\rm{sinc}}\left( \Delta\psi_n\right)}$ has a large value, which brings the large power for ${\tilde{h}_n}$. Similarly, when the practical spatial direction $\psi_l$ for the $l$th path is far away from the predefined spatial direction ${\bar \psi_n}$, ${{\rm{sinc}}\left( \Delta\psi_n\right)}$ has a small value, which brings the small power for ${\tilde{h}_n}$. It is due to the random of the practical spatial direction $\psi_l$ that the different ${\tilde{h}_n}$ can be regarded as the different Gaussian component. So, the Gaussian mixture distribution is expected to model the distribution of the beamspace channel elements.

It is worth noting that when the mean and variance of a Gaussian component are both zero, the probability density function of Gaussian distribution will be changed to
\begin{equation}\label{eq22}
{\cal C}{\cal N}\left( {\tilde{h}};0,0 \right) = \delta \left( {\tilde{h}} \right),
\end{equation}
where the $\delta \left( {\tilde{h}} \right)$ is the Dirac delta function, which means the random variable ${\tilde{h}}$ will be exact zero. Thus, the Gaussian mixture distribution can also describe the sparsity of the beamspace channel as a special case.

Then, we can derive the scalar version $\eta _{\rm{gm}}$: ${\mathbb{C} \rightarrow \mathbb{C}}$ of element-wise Gaussian mixture shrinkage function based on the the Bayesian minimize mean square error (MMSE) estimation principle~\cite{AMP1} as follows:
\begin{equation}\label{eq23}
{\eta _{\rm{gm}}} = \mathbb{E}\left\{ {{\tilde{h}}\left| {r;{\bm{\theta}} ,\sigma^2} \right.} \right\}  = \frac{\displaystyle{\int {{\tilde{h}}p\left( {r\left| {\tilde{h}} \right.;\sigma^2} \right)p\left( {{\tilde{h}};{\bm{\theta}} } \right)d{\tilde{h}}} }}{{\displaystyle\int {p\left( {r\left| {\tilde{h}} \right.;\sigma ^2} \right)p\left( {{\tilde{h}};{\bm{\theta}} } \right)d{\tilde{h}}} }},
\end{equation}
where the input element $r$ of the shrinkage function is modeled by~\cite{LAMP}
\begin{equation}\label{eq24}
r={\tilde{h}}+n,
\end{equation}
where $n$ is the additive Gaussian noise following ${\cal C}{\cal N}\left( {0,\sigma^2} \right)$. Thus, we have
\begin{equation}\label{eq25}
p\left( {r\left| {\tilde{h}} \right.;{\sigma ^2}} \right) = {\cal C}{\cal N}\left( {r;{\tilde{h}},{\sigma ^2}} \right).
\end{equation}
Given $p\left( {{\tilde{h}};{{\bm{\theta}}}} \right)$ by~(\ref{eq20}), we have
\begin{equation}\label{eq26}
\begin{array}{l}
p\left( {r\left| {\tilde{h}} \right.;{\sigma ^{\rm{2}}}} \right)p\left( {{\tilde{h}};{\bm{\theta}} } \right)\\
 = {\cal C}{\cal N}\left( {r;{\tilde{h}},{\sigma ^{\rm{2}}}} \right)\sum\limits_{k = 0}^{{N_c-1}} {{p_k}{\cal C}{\cal N}\left( {{\tilde{h}};\mu_k,\sigma _k^2} \right)} \\
 = \sum\limits_{k = 0}^{{N_c-1}} {{p_k}{\cal C}{\cal N}\left( {r;{\tilde{h}},{\sigma ^2}} \right){\cal C}{\cal N}\left( {{\tilde{h}};\mu_k,\sigma _k^2} \right)} \\
 = \sum\limits_{k = 0}^{{N_c-1}} {{p_k}{\cal C}{\cal N}\left( {r;\mu_k,{\sigma ^2} + \sigma _k^2} \right){\cal C}{\cal N}\left( {{\tilde{h}};{{\tilde \mu }_k}\left( r \right),\tilde \sigma _k^2} \right)},
\end{array}
\end{equation}
where ${\tilde \mu _k}\left( r \right) = \frac{{\sigma^2\mu_k}+{\sigma _k^2r}}{{{\sigma ^2} + \sigma _k^2}}$ and ${\tilde \sigma _k^{2}}\left( r \right) = \frac{{{\sigma ^2}\sigma _k^2}}{{{\sigma ^2} + \sigma _k^2}}$.

Finally, by  substituting~(\ref{eq26}) in~(\ref{eq23}), we can derive a new shrinkage function based on the Gaussian mixture distribution as:
\begin{equation}\label{eq27}
{\eta _{\rm{gm}}}\left( {r;{\bm{\theta}} ,{\sigma ^2}} \right) = \frac{{\sum\limits_{k = 0}^{N_c-1} {{p_k}{{\tilde \mu }_k}\left( r \right){\cal C}{\cal N}\left( {r;\mu_k,{\sigma ^2} + \sigma _k^2} \right)} }}{{\sum\limits_{k = 0}^{{N_c-1}} {{p_k}{\cal C}{\cal N}\left( {r;\mu_k,{\sigma ^2} + \sigma _k^2} \right)} }},
\end{equation}where a set of all distribution parameters ${\bm{\theta}}$ can also be called as the shrinkage parameters. Compared with the general soft threshold shrinkage function ${\bm{\eta} _{\rm{st}}}$ in the existing LAMP network, the Gaussian mixture shrinkage function ${\bm{\eta} _{\rm{gm}}}$ considering the prior distribution of the beamspace channel is designed for the specific beamspace channel estimation problem.

Now we have derived the Gaussian mixture shrinkage function, based on which we will propose the GM-LAMP network for the beamspace channel estimation in the next subsection.

\subsection{Proposed GM-LAMP network}\label{S3.2}
In order to estimate the beamspace channel more accurately, we integrate the LAMP network and the new shrinkage function derived from the Gaussian mixture distribution to propose a prior-aided GM-LAMP network.

Specifically, we replace the original soft threshold shrinkage function in the existing LAMP network by the Gaussian mixture shrinkage function. Therefore, the proposed GM-LAMP network is still constructed on the AMP algorithm. Similar to Fig. 2, the GM-LAMP network also have $T$ homogeneous layers, where the inputs and outputs of each layer are the same as those of the LAMP network. The inputs of the $t$th layer are represented by ${\bf{y}} \in {\mathbb{C}^M}$, ${{\hat{\tilde{\bf{{h}}}}}_{t}} \in {\mathbb{C}^N}$ and ${\bf{v}}_{t} \in {\mathbb{C}^M}$, where ${\bf{y}}$ is the measurement vector, ${\hat{\tilde{\bf{{h}}}}}_{t}$ and ${\bf{v}}_{t}$ are the outputs of the $\left(t-1\right)$th layer. The outputs of the $t$th layer can be represented by ${\hat{\tilde{\bf{{h}}}}}_{t+1}$ and ${\bf{v}}_{t+1}$, representing the estimate vector and residual vector of the $t$th layer, respectively. The difference is that the soft threshold shrinkage function ${\bm{\eta} _{\rm{st}}}$ of each layer is replaced by the Gaussian mixture shrinkage function ${\bm{\eta} _{\rm{gm}}}$. To do this, the channel estimate ${{\hat{\tilde{\bf{{h}}}}}_{t+1}}$ of the $t$th layer in the GM-LAMP network can be obtained by
\begin{equation}\label{eq28}
{\bf{r}}_{t}={{\hat{\tilde{\bf{{h}}}}}_{t}} + {{\bf{B}}_t}{{\bf{v}}_t},
\end{equation}
\begin{equation}\label{eq29}
{{\hat{\tilde{\bf{{h}}}}}_{t+1}} = {{\bm{\eta}}_{\rm{gm}}}\left({\bf{r}}_{t};{{\bm{\theta}}_t},{\sigma}^2\right),
\end{equation}
where ${\sigma}^2=\frac{\left\| {{{\bf{v}}_t}} \right\|_2^2}{M}$ is obtained in the same way as the AMP algorithm and the LAMP network~\cite{LAMP}, the linear transform coefficients ${{\bf{B}}_t}$ and nonlinear shrinkage parameters ${\bm{\theta}}_t$ are  trainable variables to be optimized in the training phase.

Next, we discuss how the GM-LAMP network works for the beamspace channel estimation problem in mmWave massive MIMO systems. Like most existing DNNs~\cite{DDL,ADL,CDL}, the GM-LAMP network mainly works in two phases: offline training phase and online estimation phase. In the offline training phase, given a large number of known training data, the GM-LAMP network aims to optimize overall trainable variables ${\bm{\Omega}}_{T-1}  = \{ {{\bf{B}}_t},{{\bm{\theta}} _t}\} _{t = 0}^{T-1}$ by minimizing the loss function. In the online estimation phase, by inputting the new measurements $\bf{y}$, the trained GM-LAMP network can output the estimated beamspace channel $\hat{\tilde{\bf{h}}}$. Next, we introduce these two phases in detail.
\subsubsection{Offline training phase}\label{S3.2.1}
In this paper, we adopt the supervised learning to train the GM-LAMP network. The training dataset can be represented $\{ {{\bf{y}}^d},{{\tilde{\bf{{h}}}}^d}\} _{d = 1}^D$, where ${{\bf{y}}^d}$ is the input of the GM-LAMP network, ${{\tilde{\bf{{h}}}}^d}$ is the corresponding label, and $D$ represents the number of the training data. In order to avoid overfitting, the layer-by-layer training method adopted by~\cite{LAMP} is used to train the GM-LAMP network. Generally speaking, the layer-by-layer training method can be explained from three steps.

Firstly, the whole training procedure can be divided into $T$ sequential training sub-procedures~\cite{LAMP}. For the $t$th training sub-procedure, we aim to refine the trainable variables ${\bm{\Omega}}_t  = \{ {{\bf{B}}_i},{{\bm{\theta}}_i}\} _{i = 0}^{t}$ of the $i=0$, $\cdots$, $i=t$th layer. Each layer of the GM-LAMP network has its own loss functions.

Secondly, we define two types of loss functions as follows, which are related to the linear transform operation and the nonlinear shrinkage operation:
\begin{equation}\label{eq30}
{L_t^{\rm{linear}}}\left({\bm{\Omega}}_t \right) = \frac{1}{D}\sum\limits_{d = 1}^D {\left\| {{{\bf{r}}}^d_t}({{\bf{y}}^d},{\bm{\Omega}}_t  ) - {{\tilde{\bf{{h}}}}^d} \right\|}_2^2,
\end{equation}
\begin{equation}\label{eq31}
{L_t^{\rm{nonlinear}}}\left({\bm{\Omega}}_t \right) = \frac{1}{D}\sum\limits_{d = 1}^D {\left\| {{\hat{\tilde{\bf{{h}}}}}^d_{t+1}}({{\bf{y}}^d},{\bm{\Omega}}_t  ) - {{\tilde{\bf{{h}}}}^d} \right\|}_2^2,
\end{equation}
where ${{{\bf{r}}}^d_t}$ is the output of the linear transform operation in~(\ref{eq28}), and ${{\hat{\tilde{\bf{{h}}}}}^d_{t+1}}$ is the output of the nonlinear shrinkage operation in~(\ref{eq29}) (i.e., the estimated channel of the $t$th layer). Based on these two loss functions, the training sub-procedure for the $t$th layer are further divided into the two parts: the linear training for aiming to minimizing $L_t^{\rm{linear}}$ and the nonlinear training for aiming to minimizing ${L_t^{\rm{nonlinear}}}$.

Thirdly, the hybrid method of ``individual" and ``joint" optimization is further adopted in linear training and the nonlinear training~\cite{LAMP}. Specifically, in the linear training of the $t$th training sub-procedure, only the linear transform coefficients ${{\bf{B}}_t}$ are first optimized individually, and all trainable variables ${\bm{\Omega}}_{t-1}$ of the previous $i=0$, $\cdots$, $i=(t-1)$th layer together with ${{\bf{B}}_t}$ are optimized jointly. Similarly, in the nonlinear training of the $t$th training sub-procedure, the nonlinear shrinkage parameters ${\bm{\theta}}_t$ are first optimized individually, and then all trainable variables ${\bm{\Omega}}_{t-1}$ of the previous $i=0$, $\cdots$, $i=(t-1)$th layer together with ${{\bf{B}}_t}$ and ${\bm{\theta}}_t$ are optimized jointly. Based on the above three steps, the trained GM-LAMP network can be efficiently fine-tuned in each layer, and therefore avoid bad local optimum caused by overfitting~\cite{LAMP}.
\begin{algorithm}[h]
\caption{Layer-by-Layer Training Method}
\textbf{Initialization}: ${{\bf{B}}_0} = {\bm{A}}^T, {\bm{\theta}}_0={\bm{\theta}}^0$.
\\1. \hspace*{+0.5mm} Learn ${{\bf{B}}_0}$ to minimize ${L_0^{\rm{linear}}}$
\\2. \hspace*{+0.5mm} Learn ${\bm{\theta}}_0$ with fixed  ${{\bf{B}}_0}$ to minimize ${L_0^{\rm{nonlinear}}}$
\\3. \hspace*{+0.5mm} Re-learn ${\bm{\Omega}}_0=\{{{\bf{B}}_0},{\bm{\theta}}_0\}$ to minimize ${L_0^{\rm{nonlinear}}}$
\\\textbf{for} $t = 1,\cdots,T-1$ \textbf{do}
\\4. \hspace*{+0.5mm} {Initialization}: ${{\bf{B}}_t} = {{\bf{B}}_{t-1}}, {\bm{\theta}}_t={\bm{\theta}}_{t-1}$
\\5. \hspace*{+0.5mm} Learn ${{\bf{B}}_t}$ with fixed ${{\bm{\Omega}}_{t-1}}$ to minimize ${L_t^{\rm{linear}}}$
\\6. \hspace*{+0.5mm} Re-learn $\{{\bm{\Omega}}_{t-1},{\bf{B}}_t\}$ to minimize ${L_t^{\rm{linear}}}$
\\7. \hspace*{+0.5mm} Learn ${\bm{\theta}}_t$ with fixed $\{{\bm{\Omega}}_{t-1},{\bf{B}}_t\}$ to minimize ${L_t^{\rm{nonlinear}}}$
\\8. \hspace*{+0.5mm} Re-learn ${{\bm{\Omega}}_t}=\{{\bm{\Omega}}_{t-1},{\bf{B}}_t,{\bm{\theta}}_t\}$ to minimize ${L_t^{\rm{nonlinear}}}$
\\\textbf{end for}
\\\KwOut{${{\bm{\Omega}}_{T-1}}$.}
\end{algorithm}

\textbf{Algorithm 2} shows the specific layer-by-layer training method. Steps 1-3 represent the training sub-procedure for the initial layer (i.e., $t=0$), where ${\bf{B}}_0$ and ${\bm{\theta}}_0$ are first optimized individually and then optimized jointly. Then, the training sub-procedure for the $t=1$, $t=2$, $\cdots$, $t=(T-1)$th layer are performed sequentially. Before training, trainable variables of the $t$th layer are initialized as the values for those of the $(t-1)$th layer, as shown in Step 4. Steps 5-6 and Steps 7-8 represent the linear training and the nonlinear training of the training sub-procedure for the $t$th layer, respectively. Step 5 represents the individual optimization of linear transform coefficients ${{\bf{B}}_t}$, while Step 6 represents the joint optimization of ${{\bm{\Omega}}_{t-1}}$ and ${{\bf{B}}_t}$. Similarly, Step 7 represents the individual optimization of nonlinear shrinkage parameters ${\bm{\theta}}_t$, while Step 8 represents the joint optimization of ${{\bm{\Omega}}_{t-1}}$, ${{\bf{B}}_t}$ and ${\bm{\theta}}_t$.

After overall trainable variables ${{\bm{\Omega}}_{T-1}}$ of $T$ layers are optimized, we can obtain a trained GM-LAMP network to directly estimate the beamspace channel.

\subsubsection{Online estimation phase}\label{S3.2.2}
In this phase, we apply the trained GM-LAMP network to the beamspace channel estimation problem in mmWave massive MIMO systems, where the new measurements are fed into the trained GM-LAMP network to directly generate the corresponding estimates.

Finally, the normalized mean square error (NMSE) is used to evaluate the performance of the GM-LAMP network:
\begin{equation}\label{eq32}
{\rm {NMSE}} = \frac{\mathbb{E}\left\{{\sum\limits_{k = 1}^K{\left\|{{{\hat{\tilde{\bf{h}}}}}}_k - {{\tilde{\bf{h}}}}_k \right\|_2^2}}\right\}}{\mathbb{E}\left\{{{\sum\limits_{k = 1}^K {\left\| {{\tilde{\bf{h}}_k}} \right\|_2^2}}}\right\}}.
\end{equation}

\subsection{Insights from the proposed GM-LAMP network}\label{S3.3}
From the discussion above, we can find that in the existing LAMP network~\cite{LAMP}, the soft threshold shrinkage function only utilizes the sparsity of the signal to be recovered. By contrast, the Gaussian mixture shrinkage function in the proposed GM-LAMP network is derived from the Gaussian mixture distribution, which can approximate the distribution of beamspace channel elements more accurately. With the help of more prior information, the GM-LAMP network is more suitable for the beamspace channel estimation problem.

In this paper, we refine the existing LAMP network based on the prior distribution of the beamspace channel to improve the estimation accuracy. This idea can be extended to solve other sparse signal recovery problems in wireless communications with improved performances. If we know the prior distribution of sparse signals, e.g., the sparse active users in massive machine-type communications, the sparse active antennas in spatial modulation systems, and the sparse interfering BSs in ultra-dense networks~\cite{gao18CS}, we can obtain a new shrinkage function based on the new distribution for the DNN, thus the performance can be improved.

Moreover, most existing DNNs, such as the fully connected network, have a generality for a large number of problems, but are not optimized for the specific problems to be solved. For specific problems, by leveraging the domain knowledge (besides the signal distribution considered in this paper, e.g., other statistics like mean and variance, the inherent correlation of the signal, etc.), we can design some specialized DNNs for specific problems with better performance.

\subsection{Computational complexity analysis}
In this subsection, we provide the computational complexity analysis of the proposed GM-LAMP scheme and other existing schemes. Since both the LAMP network and the GM-LAMP network are constructed on the AMP algorithm, the computational complexity of the AMP algorithm, the LAMP network and the GM-LAMP network is the same, i.e., $\mathcal{O}(TMN)$. By contrast, the computational complexity of the OMP algorithm can be represented by $\mathcal{O}(SMN)+\mathcal{O}(S^3M)$, where $S$ is the sparsity level of the beamspace channel vector~\cite{venugopal2017channel}.

\section{Simulation Results}\label{S5}
In this section, we present the beamspace channel estimation performance comparison among the proposed GM-LAMP network, the existing LAMP network, and other conventional beamspace channel estimation schemes. In order to prove the effectiveness of our work, we provide the simulation results on the widely used Saleh-Valenzuela channel model and the publicly-available DeepMIMO dataset based on ray-tracing, respectively.

\subsection{Simulation setup}\label{S4.1}
In our simulations, we consider that the BS equips a ${N=256}$ lens antenna array and ${{N_{{\rm{RF}}}}{\rm{ = 16}}}$ RF chains. The number of single-antenna users is set to ${K=16}$. The number of measurements is set to $M=128$. The SNR for uplink channel estimation is defined as ${1/\sigma _{{{n}}}^2}$. Then, we generate the spatial channel samples according to the Saleh-Valenzuela channel model and the DeepMIMO dataset.

For the the Saleh-Valenzuela channel model in~(\ref{eq1}), we set the same channel parameters for each user $k$ as follows~\cite{sayeed2013beamspace}: 1) ${L_k=3}$ path components; 2) ${\beta _{k,l} \sim {\cal C}{\cal N}\left( {0,1} \right)}$ for ${l = 1,2,3}$; 3) ${\theta _{k,l}}\sim {\cal U}\left( {-\frac{\pi }{2},\frac{\pi }{2}} \right)$, ${\theta _{k,l}^{\rm{azi}}}\sim {\cal U}\left( {-\frac{\pi }{2},\frac{\pi }{2}} \right)$ and ${\theta _{k,l}^{{\rm{ele}}}}\sim {\cal U}\left( {-\frac{\pi }{2},\frac{\pi }{2}} \right)$ for ${l = 1,2,3}$. In order to train and test the GM-LAMP network, we generate $80000$, $2000$ and $2000$ samples as the training, the validation and the testing set based on the above setup, respectively.

Next, we introduce how to obtain the channel samples using the DeepMIMO dataset. The DeepMIMO is a parameterized dataset published for deep learning applications in mmWave and massive MIMO systems, described in detail in~\cite{DeepMIMO}. The channel samples generated by the DeepMIMO are based on the ray-tracing, which can capture the dependence on key environmental factors such as the environment geometry, operating frequency and so on. One main advantage of the DeepMIMO dataset is that it is completely defined by the ray-tracing scenario and the parameters set. In our simulations, we consider the DeepMIMO dataset with the outdoor ray-tracing scenario `O1' working at the mmWave $28$ GHz and with the parameters set in Table I. Based on the parameters setup in Table I, we can generate about $54000$ channel samples between the BS 3 and the single-antenna users from the row R1000 to R1300. We split these $54000$ channel samples into three parts: $50000$ training samples, $2000$ validation samples and $2000$ testing samples.

It is noted that after obtaining spatial channel samples based on the Saleh-Valenzuela channel model and the DeepMIMO dataset, we can further generate the corresponding beamspace channel samples according to (4)-(8) and measurement samples according to (9)-(10).

\begin{table}[h]
\setlength{\abovecaptionskip}{0pt}
\setlength{\belowcaptionskip}{10pt}
\caption{The adopted DeepMIMO dataset parameters} \label{TAB1}
\vspace*{0mm}
\renewcommand\arraystretch{2}
\begin{center}
\begin{tabular}{|l|c|}
\hline \\ [-5 ex]
Parameter& Value\\
\hline \\ [-5 ex]
Active BSs & 3 \\
\hline \\ [-5 ex]
Active users & From row R1000 to row R1300 \\
\hline \\ [-5 ex]
Number of BS antennas & $(N_x,N_y,N_z)=(1,256,1);(1,16,16)$ \\
\hline \\ [-5 ex]
Antenna spacing & 0.5 \\
\hline \\ [-5 ex]
Number of paths & 3\\
\hline
\end{tabular}
\end{center}
\end{table}
\vspace*{0mm}

For the proposed GM-LAMP network and the existing LAMP network, the number of the layers is set as $T=8$, where the number of nodes for each layer is depended on the number of measurements $M$ and the dimension of the beamspace channel $N$, i.e., $M+N$. The layer-by-layer training method described in detail in Section III-B is adopted to optimize overall trainable variables with the Adam optimizer. We use a mini-batch of $128$ training samples for each updating. The training rate for individual optimization is set at $0.001$, and for the joint optimization, the training rate decreases to $0.0005$, $0.0001$ and $0.00001$ in turn when the validation errors stops decreasing. After training networks, we evaluate the performance of the trained networks on the test samples. What's more, in the GM-LAMP network, the number of the Gaussian component in the Gaussian mixture shrinkage function is set as $N_c=4$. Therefore, the nonlinear shrinkage parameters ${\bm{\theta}}_t$ of each layer $t$ have $12$ elements, which represent the probabilities, means and variances of four Gaussian components. Before the layer-by-layer training, we initialize the nonlinear shrinkage parameters ${\bm{\theta}}_0=\{0.25,0.25,0.25,0.25,0,0,0,0,0,0,0,0\}$, where the mean and variance of four Gaussian components are both set as 0 considering the sparsity of the beamspace channel. In the LAMP network, we initialize the nonlinear shrinkage parameter ${\lambda}_0=1$ following~\cite{LAMP}. For the OMP-based channel estimation scheme, we consider that the sparsity level of the beamspace channel vector is $S=24$. For the AMP-based channel estimation scheme, we set the number of iterations as $T=10$ and the empirical shrinkage parameters as $\lambda_{t}=1.1402$ for each iteration $t$, as did in~\cite{LAMP}.

\begin{figure}[htbp]
\begin{center}
\vspace*{0mm}\includegraphics[width=1\linewidth]{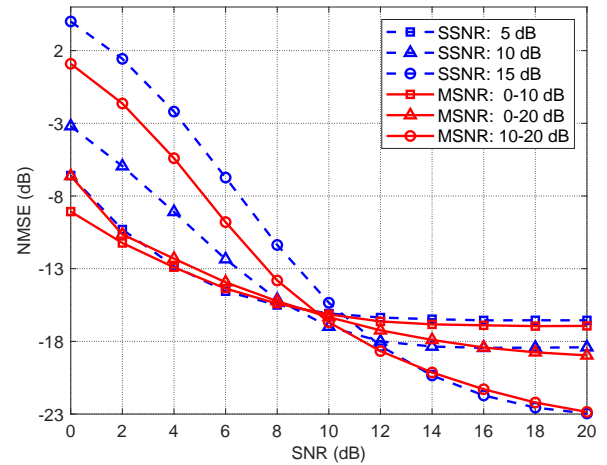}
\end{center}
\setlength{\abovecaptionskip}{-0.0cm}
\vspace*{-3mm}\caption{NMSE performance comparison of different trained GM-LAMP networks with different training settings.} \label{FIG4}
\end{figure}

The different training SNR settings have different effects on the performance of the GM-LAMP network. In order to find out what if the preferable SNR settings for the training phase, we provide the NMSE performance comparison of different trained GM-LAMP networks with different training settings, as shown in Fig. 3. Therein, we consider the Saleh-Valenzuela channel model for ULAs. The blue curves represent that all training samples are generated based on the same SNR, while the red curves represent that the training samples are generated based on multiple SNRs. For example, `SSNR: $5$ dB' means all measurement samples for training are generated when the SNR is $5$ dB. `MSNR: $0$-$10$ dB' means that the SNR (in dB) of each sample is randomly drawn from the range $[0,10]$. From Fig. 4, we can find that the trained network with the single SNR usually achieves good performance near the training SNR, but it is poor at other SNRs. In order to achieve good performance at both high SNRs and low SNRs, we train one network at SNRs between $0$ dB and $10$ dB for solving channel estimation problems at low SNRs ($0-10$ dB) and another network trained at SNRs between $10$ dB and $20$ dB for solving problems at high SNRs ($10-20$ dB) in the following simulation results.

\subsection{Simulation results on the Saleh-Valenzuela channel model}\label{S4.2}
In this subsection, we provide the beamspace channel estimation performance comparison of the OMP algorithm~\cite{alkhateeb2014channel}, the AMP algorithm~\cite{AMP1}, the LAMP network~\cite{LAMP} and the proposed GM-LAMP network based on the Saleh-Valenzuela channel model.

\begin{figure}[htbp]
\begin{center}
\vspace*{0mm}\includegraphics[width=1\linewidth]{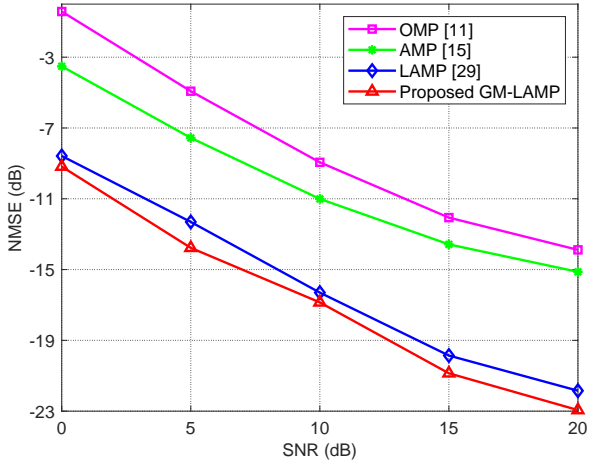}
\end{center}
\setlength{\abovecaptionskip}{-0.0cm}
\vspace*{-3mm}\caption{NMSE performance comparison for ULAs based on the Saleh-Valenzuela channel model.} \label{FIG5}
\end{figure}

Fig. 4 shows the NMSE performance comparison of four different schemes mentioned above against different SNRs, where the ULA is considered. We can observe that compared with the other three existing schemes under investigation, the proposed GM-LAMP network enjoys lower estimation errors in all considered SNR regions. In particular, we can observe that the NMSE performance of the OMP algorithm and the AMP algorithm is poor, whereas the two DL based schemes (i.e., the LAMP network and the GM-LAMP network) can achieve better NMSE performance. Moreover, thanks to considering the prior distribution of the beamspace channel, the proposed GM-LAMP network has better channel estimation accuracy than the LAMP network.

\begin{figure}[htbp]
\begin{center}
\vspace*{0mm}\includegraphics[width=1\linewidth]{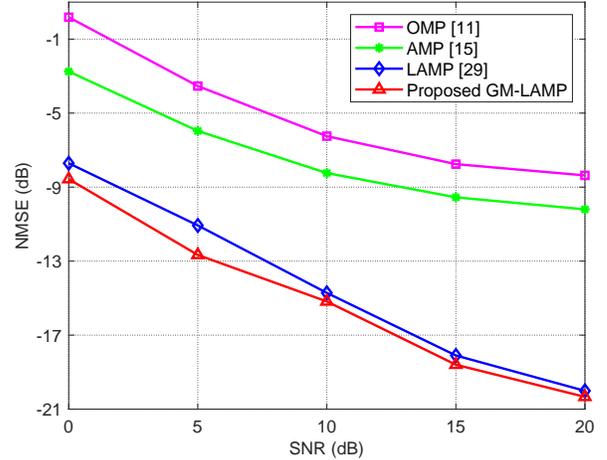}
\end{center}
\setlength{\abovecaptionskip}{-0.0cm}
\vspace*{-3mm}\caption{NMSE performance comparison for UPAs based on the Saleh-Valenzuela channel model.} \label{FIG6}
\end{figure}

In Fig. 5, we further compare the NMSE performance of four different schemes for the $16\times 16$ UPA. We can observe that the conventional OMP algorithm and AMP algorithm cannot achieve satisfactory estimation accuracy. By contrast, the proposed GM-LAMP network can still outperform the other three schemes when the antenna array is UPA.

\subsection{Simulation results on the DeepMIMO dataset}\label{S4.3}
In this subsection, we provide the beamspace channel estimation performance comparison of the proposed GM-LAMP network and the other three existing schemes based on the DeepMIMO dataset.

\begin{figure}[htbp]
\begin{center}
\vspace*{0mm}\includegraphics[width=1\linewidth]{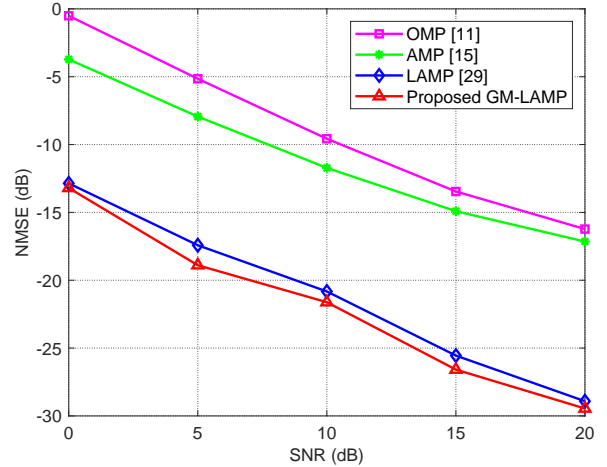}
\end{center}
\setlength{\abovecaptionskip}{-0.0cm}
\vspace*{-3mm}\caption{NMSE performance comparison for ULAs based on the DeepMIMO dataset.} \label{FIG7}
\end{figure}

\begin{figure}[htbp]
\begin{center}
\vspace*{0mm}\includegraphics[width=1\linewidth]{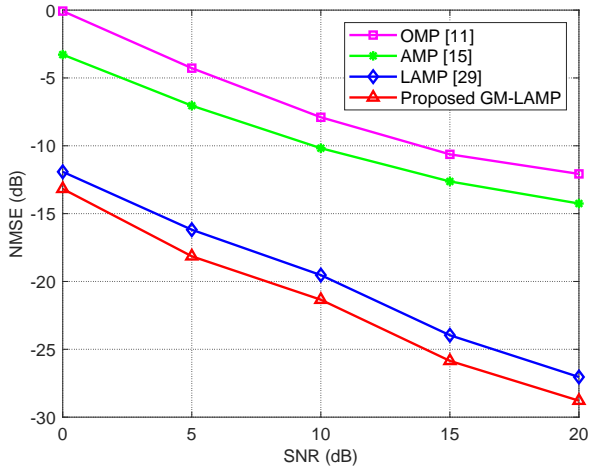}
\end{center}
\setlength{\abovecaptionskip}{-0.0cm}
\vspace*{-3mm}\caption{NMSE performance comparison for UPAs based on the DeepMIMO dataset.} \label{FIG8}
\end{figure}

Fig. 6 and Fig. 7 respectively show the NMSE performance comparison of four different schemes against different SNRs for the $256\times 1$ ULA and the $16\times16$ UPA. We can observe that the proposed GM-LAMP network can still achieve better beamspace channel estimation accuracy based on the more practical DeepMIMO dataset.

\subsection{Other simulation results}\label{S4.4}
Based on the above setup that the number of antennas at the BS is $N=256$ and the number of measurements is $M=128$, the number of complex multiplications required by four different algorithms can be calculated. The OMP algorithm requires about $2.9 \times 10^6$ complex multiplications, and the AMP algorithm requires about $6.6 \times 10^5$ complex multiplications. By contrast, the LAMP network and the GM-LAMP network require about $5.3 \times 10^5$ and $6.1 \times 10^5$ complex multiplications, respectively. In order to show the computational complexity of the proposed scheme and the existing schemes more clearly, we provide the number of complex multiplications comparison against the number of antennas $N$ at the BS, as shown in Fig. 8. It is noted that the number of measurements is set to $M=N/2$.
\begin{figure}[htbp]
\begin{center}
\vspace*{0mm}\includegraphics[width=1\linewidth]{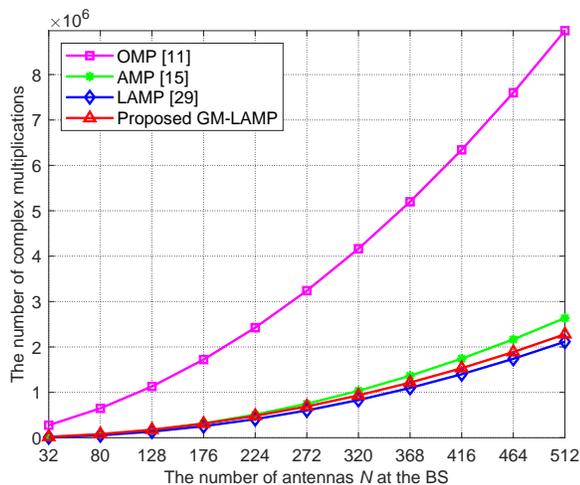}
\end{center}
\setlength{\abovecaptionskip}{-0.0cm}
\vspace*{-3mm}\caption{The number of complex multiplications against the number of antennas $N$.} \label{FIG9}
\end{figure}

From Fig. 8, we can find that the conventional OMP algorithm requires more complex multiplications than AMP-type algorithms. What's more, thanks to the powerful learning of DNNs, the LAMP network and the GM-LAMP network can converge faster than the AMP algorithm. Therefore, the number of complex multiplications required by the LAMP network and the GM-LAMP network is smaller than that required by the AMP algorithm~\cite{LAMP}. It is noted that the proposed GM-LAMP network employs a more complex Gaussian mixture shrinkage function considering the prior distribution of beamspace channel elements, which slightly increases the number of complex multiplications but achieves better estimation performance compared with the existing LAMP network.

\begin{figure}[htbp]
\begin{center}
\vspace*{0mm}\includegraphics[width=1\linewidth]{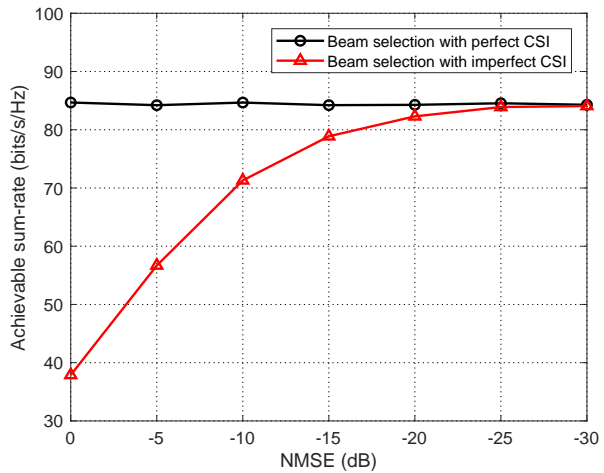}
\end{center}
\setlength{\abovecaptionskip}{-0.0cm}
\vspace*{-3mm}\caption{Sum-rate for beam selection against different NMSE for the beamspace channel estimation.} \label{FIG10}
\end{figure}

Next, we will evaluate the impact of the NMSE for the beamspace channel estimation on beam selection. In this paper, we adopt the interference-aware (IA) beam selection scheme proposed in~\cite{gao16bs}, where the downlink SNR for beam selection is defined as ${1/\sigma _{{{d}}}^2}$ with $\sigma _{{{d}}}^2$ representing the power of the receiving noise at the user side. Fig. 9 shows the sum-rate achieved by the IA beam selection against the NMSE for the beamspace channel estimation. In our simulations, we follow~\cite{Hybrid} to model the estimated beamspace channel (imperfect CSI) as
\begin{equation}\label{eq33}
{\hat{\tilde{\bf{H}}}}={\tilde{\bf{H}}}+ {\bf{E}},
\end{equation}
where ${\tilde{\bf{H}}}=[{\tilde{\bf{h}}}_1,{\tilde{\bf{h}}}_2,\cdots,{\tilde{\bf{h}}}_K]$ represents perfect CSI for $K$ users, ${\bf{E}}$ is the error matrix with entries following the distribution independent and identically distributed (i.i.d.) $\mathcal{CN}(0, {\rm{NMSE}})$. Besides, we consider the IA beam selection with perfect CSI as our benchmark.

In Fig. 9, the ULA based on the Saleh-Valenzuela channel model is considered. We provide the sum-rate comparison achieved by the IA beam selection between imperfect CSI and perfect CSI with the downlink SNR of $10$ dB. From Fig. 4, we can observe that the proposed GM-LAMP network can achieve an NMSE of about $- 23$ dB with the reduced pilot overhead by half at the SNR of $20$ dB. Fig. 9 shows that compared with the case with perfect CSI, the sum-rate loss caused by the beam selection with imperfect CSI is less than $5\%$ when NMSE is about $- 23$ dB.

\begin{figure}[htbp]
\begin{center}
\vspace*{0mm}\includegraphics[width=1\linewidth]{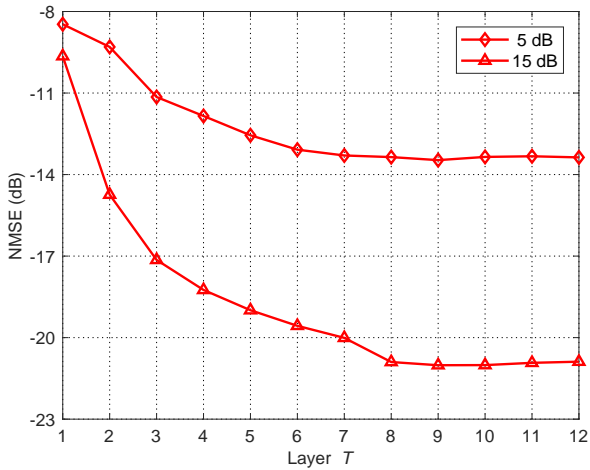}
\end{center}
\setlength{\abovecaptionskip}{-0.0cm}
\vspace*{-3mm}\caption{NMSE performance against the number of layers for the GM-LAMP network.} \label{FIG11}
\end{figure}

In order to show the convergence of the proposed GM-LAMP network, we further provide the simulation result about the NMSE performance against the number of layers. Here, we also consider the ULA based on the Saleh-Valenzuela channel model. The convergence results with different SNRs are presented in Fig. 10, which shows that the GM-LAMP network can reach convergence about at layer T = 8.

It is noted that since the orthogonal pilot transmission strategy is adopted among multiple users, pilots from different users can be distinguished without any inter-user interferences. Consequently, the NMSE performance for the beamspace channel estimation has nothing to do with the number of users.

\section{Conclusions}\label{S6}
In this paper, we have proposed a prior-aided GM-LAMP network to solve the beamspace channel estimation problem in mmWave massive MIMO systems. Specifically, we first derive a new shrinkage function by exploiting the Gaussian mixture prior distribution of beamspace channel elements. Different from the original shrinkage function in the existing LAMP network, the derived Gaussian mixture shrinkage function can embody more prior information of the beamspace channel besides sparsity. Then, by integrating the LAMP network with the Gaussian mixture shrinkage function, a GM-LAMP based beamspace channel estimation scheme is developed. To verify the performance of our work, we provide simulation results on the Saleh-Valenzuela channel model and the ray-tracing based DeepMIMO dataset, respectively. Simulation results show that compared with the existing LAMP network and other conventional beamspace channel estimation schemes, the proposed GM-LAMP network considering the prior distribution can achieve better estimation accuracy with a low pilot overhead. We can find by leveraging the domain knowledge of the problems to be solved, the general DNN can be redesigned to improve the performance for the specific problems. For future work, we will follow the idea of the proposed GM-LAMP network to solve the channel estimation problem in terahertz (THz) communications by considering THz channel features.

\bibliography{IEEEabrv,GM_LAMP}

\end{document}